# Composition induced metal–insulator quantum phase transition in the Heusler type Fe$_2$VAl

Takashi Naka,[1,*] Artem M. Nikitin,[2] Yu Pan,[2] Anne de Visser,[2] Takayuki Nakane,[1] Fumihiro Ishikawa,[3] Yuh Yamada,[3] Motoharu Imai,[1] and Akiyuki Matsushita[1]

[1]*National Institute for Materials Science, 1-2-1 Sengen, Tsukuba, Ibaraki 305-0047, Japan*
[2]*Van der Waals–Zeeman Institute, University of Amsterdam, Science Park 904, 1098 XH Amsterdam, The Netherlands*
[3]*Department of Physics, Faculty of Science, Niigata University 2-8050, Ikarashi, Niigata 950-2181, Japan*

We report the magnetism and transport properties of the Heusler compound Fe$_{2+x}$V$_{1-x}$Al at $-0.10 \leq x \leq 0.20$ under pressure and a magnetic field. A metal–insulator quantum phase transition occurred at $x \approx -0.05$. Application of pressure or a magnetic field facilitated the emergence of finite zero-temperature conductivity σ$_0$ around the critical point, which scaled approximately according to the power law $(P - P_c)^\gamma$. At $x \leq -0.05$, a localized paramagnetic spin appeared, whereas above the ferromagnetic quantum critical point at $x \approx 0.05$, itinerant ferromagnetism was established. At the quantum critical points at $x = -0.05$ and 0.05, the resistivity and specific heat exhibited singularities characteristic of a Griffiths phase appearing as an inhomogeneous electronic state.

*Corresponding author.
Tel.: +81-29-859-2730; Fax: +81-29-859-2701; E-mail: NAKA.Takashi@nims.go.jp

# 1. Introduction

The intriguing, complicated magnetic phase diagrams of itinerant magnets and Kondo lattice systems reveal the existence of exotic phenomena related to magnetic instabilities, which are in many cases associated with the proximity of a quantum critical point (QCP), where the magnetic transition temperature vanishes (that is, $T_{C,N} \to 0$) or the magnetic ground-state transition takes place at $T = 0$. In Kondo lattice systems, the Ruderman-Kittel-Kasuya-Yosida couplings between the 4f or 5f spins mediated by conduction band spins compete with the Kondo coupling, thereby disrupting any magnetic orderings. The magnetic transition temperature can be tuned (and driven to zero) by sweeping a control parameter, such as the magnetic field, external mechanical pressure, or element composition. Since the 1970s, it has often been observed that in the vicinity of the QCP, Fermi liquid behavior breaks down, and singularities in various thermodynamic quantities, such as specific heat and magnetic susceptibility, arise [1]. More recently, high-$T_c$ cuprate and heavy fermion systems exhibiting exotic superconductivities have been discovered and are investigated extensively. These systems continue to be an active area of research in condensed matter physics. These investigations have indicated that exotic superconductivities often emerge in the vicinity of a magnetic QCP. It is important to consider how the distinctive behaviors of physical properties around a QCP, referred to as quantum critical singularities, are affected by the introduction of disorder in a crystal structure by, for example, elemental substitution [2]. In the case of materials that show a metal–insulator (MI) transition, the effect of disorder on quantum critical singularities must be considered because the kinetic energy of conduction electrons $\varepsilon_F$ is comparable to energy fluctuations by disorder, electron–impurity coupling, and electron–electron interactions. This competitive situation brings about fascinating new complex phenomena. Practical examples of complex phenomena have been reported [2, 3], such as the inhomogeneous states observed in (La,Sr)MnO$_3$, which features an enormous number of low-lying metastable states due to the presence of disorder and impurities. Generally, crystallographic disorder, which is often equivalent to charge distribution disorder or inhomogeneity, competes with Coulomb interactions between electrons, which, in contrast to crystallographic disorder, preferably stabilize periodic charge distribution. Similarly, disorder seems to reduce magnetic transition temperatures.

The Heusler compound Fe$_2$VAl is a nonmagnetic semimetal [4–6]. Induction of off-stoichiometry and antisite defects by means of thermal and mechanical treatments can be used to modify the physical properties of Fe$_2$VAl, particularly its magnetic ground state [7–9]. Graf *et al.* pointed out that Fe$_2$VM (M = Al or Ga), like Fe$_2$TiSn, is located at a nonmagnetic node on the Slater–Pauling curve of spontaneous ferromagnetic moment $M_t$, represented as $M_t = Z - 24$, where $Z$ is the number of valence electrons [10]. This $M_t - Z$ relation has been confirmed experimentally for Heusler compounds. Note, however, that ferromagnetism can be induced by crystallographic defects, such as antisite defects, even in stoichiometric Fe$_2$VAl [7, 8]. If an Fe atom is located at the V site, surrounded by four Fe atoms at the Fe sites, the relatively large magnetic moment of this antisite Fe (theoretically estimated to be 2.2$\mu_B$ [4] and 2.3$\mu_B$ [5]) is stabilized and induces moments (approximately 1 $\mu_B$/Fe) of the surrounding Fe atoms [4, 5]. These theoretically estimated values seem to be consistent with many experimental findings in Fe$_2$VAl. It has been speculated that in an off-stoichiometric system, Fe$_{2+x}$V$_{1-x}$Al, a sharp MI transition occurs at $x < 0$ [11, 12]. At $x = -0.02$, that is, in the paramagnetic region, resistivity shows metallic conductivity below $T = 240$ K [8, 9, 13], whereas at $x = -0.05$, variable-range hopping conduction is observed at $2 < T < 40$ K [14, 15]. As shown in the $x$–$T$ magnetic phase diagram presented by Naka *et al.* [12], the ferromagnetic QCP $x_c^m$ is at $x = +0.05$, and above this point, $T_C$ and a spontaneous moment arise and then increase linearly with increasing $x$. It should be emphasized that many experimental and theoretical investigations of off-stoichiometric systems have focused on the boundary between the paramagnetic and ferromagnetic states. $Fe_{2+x}V_{1-x}Al$ and $Fe_2VAl_{1-\delta}$ at $x < 0$ and $\delta > 0$, respectively, exhibit semiconducting behavior, which suggests the presence of a MI transition resulting in the observed singularities and enhancements of thermodynamic quantities [12]. In this paper, which focuses mainly on the $x$ variation of the electronic states in $Fe_{2+x}V_{1-x}Al$ at $x < 0$, we present comprehensive measurements of macroscopic quantities, magnetoresistance, Hall coefficient, specific heat, and resistivity under an applied magnetic field and under high pressure, and our results reveal that a MI transition takes place at $x \approx -0.05$. Experimental and theoretical studies of Fe-based Heusler compounds in which the number of valence electrons $Z$ is reduced from 24 have been reported for $Fe_2V_{1-x}Ti_xAl$ ($0 \leq x \leq 1$) [16] and $Fe_{3-x}V_xAl$ ($0 \leq x \leq 3$) [17]. The former exhibits a ferromagnetic transition at $x \approx 0.1$ with itinerant magnetic characteristics at $x > 0$. In contrast, in this study, we confirmed that in $Fe_{2+x}V_{1-x}Al$, a paramagnetic insulating state appears at $x < -0.05$. We interpret the negative temperature coefficient of resistivity (TCR) and the singular temperature variation in the specific heat that are observed as ubiquitous characteristics at lower temperatures, as being due to an electron–electron correlation of conduction band electrons and the distribution of a characteristic energy, such as the Kondo temperature $T_K$. Finally, we present an $x$–$T$ phase diagram for $Fe_{2+x}V_{1-x}Al$ with $-0.10 \leq x \leq 0.20$ covering the MI and the paramagnetic–ferromagnetic phase boundaries.

## 2. Experimental methods

Polycrystalline samples of $Fe_{2+x}V_{1-x}Al$ with $-0.10 \leq x \leq 0.20$ were prepared by means of an arc melting method. After the amounts of elements needed to achieve the desired nominal chemical composition were arc melted several times, the specimens were sealed in a vacuum quartz tube, homogenized by heating at 1000 °C for 15 h, and then annealed at 400 °C for 15 h. To investigate the dependence of the insulator phase formation on sample preparation, we examined a postannealed sample (sample 1) and an as-cast sample (sample 2) with $x = -0.10$.

X-ray diffraction measurements revealed that all of the samples had the Heusler-type ($L2_1$) structure without any secondary phases. Because the vapor pressure of aluminum is higher than the vapor pressures of Fe and V, a small amount of aluminum (a few percent) can be expected to evaporate during arc melting. Therefore, the true chemical formula of the obtained alloys can be expressed as $Fe_{2+x}V_{1-x}Al_{1-\delta}$ with $\delta \approx 0.03$, but for convenience, we use the starting composition in referring to the samples hereafter.

DC magnetization was measured down to $T = 1.9$ K with a conventional superconducting quantum interference device magnetometer (MPMS-XL, Quantum Design). Specific heat and magnetoresistance were measured down to $T = 2$ K with a physical property measurement system (Quantum Design). Low-temperature resistivity was measured down to $T = 0.23$ K in a $^3$He refrigerator (Heliox VL, Oxford Instruments) with a sensitive LR700 (Linear Research) AC resistance bridge at the Van der Waals–Zeeman Institute of the University of Amsterdam. Resistivity under high pressure (up to 2.2 GPa) was measured with a hybrid clamp cell made of CuBe and NiCrAl alloys with the pressure-transmitting liquid media Daphne 7373 (Idemitsu).

## 3. Results and discussion

*3.1 Magnetization*

Quantitative magnetic investigation of specimens of $Fe_{2+x}V_{1-x}Al$ in the paramagnetic state is difficult because a ferromagnetic component dominates the magnetic responses, especially at low magnetic fields. Therefore, we carried out comprehensive magnetic measurements for Fe$_{2+x}$V$_{1-x}$Al with $-0.10 \leq x \leq 0.05$. Magnetization $M$ as a function of magnetic field $H$ measured at various temperatures can be represented as $M(x, H, T) = m_0(x, H, T) + \chi(x, T)H$, where $m_0$ and $\chi$ are residual magnetization and magnetic susceptibility, respectively (Fig. 1(a)). Above $H \approx 10$ kOe, $m_0$ and $\chi$ are field independent. In fact, $m_0$ and $\chi$ obtained from $MH$ curves above $H = 10$ kOe corresponded quantitatively to those obtained from $M(T)$ curves measured at $H = 20$ and 50 kOe (Fig. 1(b)). Plots of $m_0$ and $\chi$ versus $T$ confirmed that at $x = -0.10$, the residual ferromagnetic moment $m_0$ was strongly sample dependent, whereas $\chi$ was not (Fig. 1(c)). Therefore, the paramagnetic susceptibility $\chi$ seems to be intrinsic. In contrast, $m_0$ results from a ferromagnetic impurity phase or a ferromagnetic cluster, such as an Fe$_3$Al cluster generated as an antisite defect, embedded in the paramagnetic host. For the paramagnetic region of $-0.10 \leq x \leq 0.05$, the magnetic susceptibility obtained by using $M(T)$ curves measured at $H = 20$ and 50 kOe, that is, $\chi(T) = \Delta M/\Delta H$, the reciprocal susceptibility $\chi^{-1}(T)$ and the residual ferromagnetic moment $m_0(T)$ at $H = 0$ are shown in Fig. 2(a)–(c), respectively.

The magnetic parameters of the paramagnetic component (Fig. 2(a)–(b)) were obtained by means of a modified version of the Curie–Weiss law, $\chi(T) = \chi_0 + C_M/(T - \theta)$, where $\chi_0$, $C_M$, and $\theta$ are a constant susceptibility, a Curie constant, and a Weiss temperature, respectively. The effective magnetic moment $p_{eff}$, $\theta$, and $\chi_0$, are shown as a function of $x$ in Fig. 3(a)–(c), respectively. Because $\theta \approx 0$ at $x < 0$, it is plausible that magnetic interactions between electron (hole) spins were suppressed and, therefore, that the spin was localized rather strongly. In addition, $p_{eff}$ can be reproduced by the localized spin model: $p_{eff} = [3xg^2S(S + 1))]^{1/2}$, where $g$ is the $g$-value of the localized spin, because the number of electrons (holes) with $S = 1/2$ doped by the off-stoichiometry was $3x$ per unit formula of Fe$_{2+x}$V$_{1-x}$Al. As will be shown below, however, the observed value of $p_{eff}$ at $x \leq 0$ could be reproduced by using the $J = 3/2$ paramagnetic cluster model adopted by Lue *et al*. with parameters derived by specific heat measurement under magnetic field [18].

*3.2 Specific heat*

The temperature dependences of the electronic contribution to specific heat $\Delta C/T = [C(x) - C_{HT}(0)]/T$ at magnetic fields of $H = 0$ and 80 kOe are shown in Fig. 4(a)–(b), respectively, where $C_{HT}$ was estimated to fit an experimental value of $C(x = 0)$ by means of $\gamma T + \beta T^2 + \eta T^4$ above $T = 8$ K and was estimated numerically by means of a polynomial series with respect to temperature for $T > 30$ K [18, 12]. As indicated previously, the low-temperature upturn in $\Delta C/T$ [9, 11, 12, 18–21] and the Schottky-like peak at $T \approx 50$ K [12, 21] are ubiquitous in the range $-0.10 \leq x \leq 0.20$. The low-temperature upturn in $\Delta C/T$ observed at $x \leq 0.02$ approximately followed a power law with respect to temperature, $T^{\alpha-1}$, down to $T = 2$ K, with a nonuniversal exponent $\alpha$, which varied continuously as a function of $x$, specifying divergence of thermodynamic quantities (Fig. 4(a), inset). The characteristic divergences are reminiscent of a quantum Griffiths phase, which occurs in many disordered correlated systems. The existence of a quantum Griffiths phase in the vicinity of the paramagnetic–ferromagnetic QCP has been confirmed by comprehensive measurements of thermodynamic quantities of transition metallic compounds such as $Fe_{2+x}V_{1-x}Al$ [21], $FeV_{1-x}M_xGa$ (M = Ti and Rh) [22], $Ni_{1-x}V_x$ [23], and $Fe_{1-x}Co_xS_2$ [24] and in the rare earth intermetallic compound $Ce_{1-x}La_xRhIn_5$ [25].

In stoichiometric $Fe_2VAl$, the Non-Fermi-liquid like behaviors first reported by Nishino *et al.* [11] and the field-induced Schottky-like anomaly in specific heat reported by Lue *et al.* [18] are still controversial. Lue *et al.* contended that the former is due to the distribution of magnetic anisotropies and that the latter results from paramagnetic clusters. The Schottky specific heat curves measured at various magnetic fields were precisely reproduced by using a multilevel Schottky function with a spin $J$ of 3/2 and a $g$ value of 1.93 [18]. By using their analysis method, that is, by employing the multilevel Schottky function with the assumption of $J = 3/2$ for $\Delta C/T$ measured at $H = 80$ kOe in $Fe_{2+x}V_{1-x}Al$, we derived the number of Schottky centers (the number of clusters) $N_{cluster}$ and the energy gap $\Delta$ ($= g\mu_B H/k_B$) for $-0.10 \leq x \leq 0.10$ (Fig. 5(a) and (b)). The energy gap $\Delta$ for $x < 0$ agreed well with that derived by Lue *et al.* [18]. As shown in Fig. 5(b), at $x \leq 0$, $g$ was estimated to have a constant value of 1.93 within an error of $\pm 0.07$, and the calculated effective moment, $p_{eff}^{cal} = [N_{cluster} gJ(J + 1)]^{0.5}$, reproduced the observed moment $p_{eff}^{obs}(x)$ derived from the magnetic susceptibility $\chi(x, T)$. It is plausible that doped hole spins associated with a magnetic cluster behave as localized spins. Although Lue *et al.* deduced that the $J = 3/2$ spin cluster consists of antisite Fe atoms [18], we speculate, on the basis of our findings at $x < 0$, that the magnetic cluster is associated with antisite V atoms rather than with antisite Fe atoms. Note that at $x > 0$, $\Delta$ and $T_{max}(\Delta C/T)$ obtained at $H = 80$ kOe increased to values corresponding to implausibly large values of $g$ ($g = 3$–4; Fig. 5(b)) and that $p_{eff}^{cal}(x)$ deviated considerably from $p_{eff}^{obs}(x)$ (Fig. 5(a)). These inconsistencies with the $J = 3/2$ cluster model can be explained reasonably well by the fact that the antisite Fe occupying the V site facilitates the establishment of itinerant (ferromagnetic) features in $Fe_{2+x}V_{1-x}Al$ at $x > 0$ [12] and the fact that an inhomogeneous state (Griffiths phase) that appears to consist of ferromagnetic and paramagnetic clusters emerges at $0 < x < x_c^m$ [21]. In a theoretical study, Neto *et al.* found that a Schottky anomaly in the specific heat, due to a magnetic field, appears close to the QCP for magnetic ordering in magnetic metallic systems [26].

Likewise, as we previously claimed for $Fe_{2+x}V_{1-x}Al$ near the paramagnetic–ferromagnetic QCP at $x \approx 0.05$ [12, 21], the low-temperature enhancement in $\Delta C/T$ that we observed at $x \approx -0.05$ and the singular temperature dependence, approximated by $T^{\alpha-1}$, can be reminiscent of a Griffiths-McCoy singularity near the QCP (Fig. 4(a)) [26]. As indicated below, there was an obvious MI transition in the vicinity of $x = -0.05$ and $T < 0.25$ K. The exponent α(x) obtained from low-temperature specific heat data showed the characteristic features expected for a Griffiths phase [26]; that is, α was approximately 0 at the quantum critical point of $x_{MIT}$ and recovered to that of a Fermi liquid α ≈ 1, with increasing distance from the QCP (Fig. 4(a), inset). Note that α was also reduced near the paramagnetic–ferromagnetic QCP at $x_c^m$. It is plausible that at $x > 0$, the field-induced Schottky anomaly could be attributed to the ferromagnetic Griffiths singularity/phase, as we previously speculated [21]. On the other hand, at $x < 0$ we need to consider whether the individual paramagnetic cluster [18] or a Griffiths state associated with the MI quantum phase transition is responsible for the anomaly.

*3.3 Magnetoresistance and Hall coefficient*

Figure 6(a) shows the temperature dependences of the resistivity $\rho_{xx}$ at $H = 0$ and 120 kOe for $x = -0.10$ and $-0.05$. In the vicinity of the ferromagnetic QCP, the magnetoresistance ratio $\Delta\rho_{xx}/\rho_{xx}$ saturates to a constant value with decreasing temperature [19]. In contrast, we found that below $x = -0.05$, $\Delta\rho_{xx}/\rho_{xx}$ diverged with decreasing temperature (Fig. 6(b)). The value of $\Delta\rho_{xx}/\rho_{xx}(T = 0.25$ K$)$ was clearly at a minimum at $x \approx -0.05$, whereas $\Delta\rho_{xx}/\rho_{xx}(T = 4$ K$)$ was not (Fig. 7(a)). As mentioned previously [15] and later in this paper, the pressure coefficient of resistivity $(1/\rho_{xx})(d\rho_{xx}/dP)$ and $C/T$ at $T = 2$ K were at a minimum and a maximum at $x \approx -0.05$, respectively (Fig. 7(b)–(c)). These anomaly seem to reflect the presence of a MI quantum phase transition at $x \approx -0.05$.

Generally, the Hall coefficient $R_H$ provides information about the conduction carriers in a material. In magnetic materials, however, the number of carriers is difficult to obtain from the normal Hall coefficient $R_0$ because the contribution of the anomalous Hall coefficient is large compared to that of the normal Hall coefficient. The Hall coefficient $R_H$ is represented as

$$R_H = \rho_{xy}/B = R_0 + R_s M/H \qquad (1)$$

where $B$, $\rho_{xy}$, and $R_s$ are the magnetic flux density, the Hall resistivity, and the anomalous Hall coefficient, respectively. In stoichiometric $Fe_2VAl$, the numbers of the two carriers densities are difficult to obtain accurately because both electron and hole carriers contribute to conduction. In an experimental study of $Fe_{2+x}V_{1-x}M$ (M = Al or Ga) compounds, Fukuhara *et al.* determined $R_0$ and $R_s$ separately in the semimetallic and ferromagnetic metallic regions by means of comprehensive measurements of transport and magnetic properties [27]. For both compounds, the $R_0$ was of the order of $10^{-8}$ m$^3$/C, which is comparable to that for $Fe_{3-x}V_xAl_y$ measured by Matsushita *et al.* [8]. The temperature dependences of $R_H$ for $x = -0.02, -0.05$ [8], and $-0.10$ are shown in Fig. 8(a). Note that the sample dependence of $R_H(T)$ was negligibly small (compare samples 1 and 2). This agreement provides a simple basis for estimating the anomalous Hall term; that is, $M/H$ in eq. (1) can be replaced with $(M - m_0)/H$. We confirmed that above $T = 15$ K, the Hall resistivity $\rho_{xy}$ was linearly related to magnetic field up to $H = 50$ kOe. If we employ a rigid band model, it is expected that at $x < -0.02$, the number of holes can be approximately equal to $x$. Using a previously reported procedure for estimating $R_0$ [28], we tried to derive $R_0$ from the $R_H(T)$ data for $x = -0.05$ and $-0.10$. Because at $T > T_{max}$, $\rho_{xy}$ seemed to be proportional to $\rho_{xx}^\delta$ with $\delta \approx 2$ (Fig. 8(b)) for $x = -0.05$ and $-0.10$, we expected that the anomalous term consisted mainly of the side-jump contribution, $R_s \propto \rho_{xx}^2$. To estimate $R_0$, we used the following general formula:

$$R_H = R_0 + S_H' \chi \rho_{xx}^\delta \qquad (2)$$

where $S_H'$ is a constant and $\chi$ is magnetic susceptibility, which is given by $(M - m_0)/H$). The values of $R_0(n_c)$ were estimated to be $1.5 \times 10^{-9}$ (0.26/f.u.) and $1.2 \times 10^{-9}$ m$^3$/C (0.21/f.u.) for $x = -0.05$ and $-0.10$, respectively, where $n_c$ is the number of conduction carriers. As mentioned above, these values are positive in sign and agree roughly with the number of doped valence holes ($= 3x$) in Fe$_{2+x}$V$_{1-x}$Al. The values of the exponent $\delta$ were 1.6 and 2.0 for $x = -0.05$ and $-0.10$, respectively. Here we assumed that $R_0$ was temperature independent at $T \gg T_{max}$.

*3.4 Pressure- and field-induced metal–insulator transition*

Figure 9(a)–(c) shows the pressure dependences of $\rho_{xx}$ at $T = 2$ and 290 K for $x = 0.10$, $-0.02$, and $-0.10$, respectively. Resistivity can be expressed as $\rho_{xx} = \rho_{wl} + \rho_{ec} + \rho_{spin}$, where $\rho_{wl}$, $\rho_{ec}$, and $\rho_{spin}$ are components due to weak localization resulting from disorder, electron correlation, and magnetic scattering, respectively. In the ferromagnetic region ($x > x_c^m$), we detected an anomaly in $\rho_{xx}(T)$ that accompanied the ferromagnetic transition at around $T_C$ because it is generally expected that $\rho_{spin}$ decreases at $T \leq T_C$. Simply put, at the limit of $T \to 0$, $\rho_{spin}(T = 0)$ seems to be described by a stepwise function at $x = x_c^m$; that is, $\rho_{spin} = \rho_{spin}(0)\theta(x - x_c^m)$, where $\theta(x)$ is a step function with respect to $x$. From this equation, we can readily obtain the pressure coefficient, $(1/\rho_{spin})(d\rho_{spin}/dP) = -\theta^{-1}(d\theta/dx)(dx_c^m/dP)$, which displays a sharp anomaly at $x \approx x_c^m$. Note that $x_c^m$ increased with increasing pressure because the positive pressure coefficient corresponds to that of $dx_c^m/dP$ (Figs. 7(b) and 9(a)). This positive coefficient of $x_c^m$ is consistent with the fact that $T_C$ has been observed to decrease with pressure [15].

Similarly, this argument about the pressure dependence of resistivity in the vicinity of $x_c^m$ can be used to explain the pressure dependence of $\rho_{xx}$ in the vicinity of the MI transition. If we assume that $n_c \approx x$, resistivity is proportional to $(x_{MI} - x)^{-\xi}$ with $\xi > 0$. This assumption gives $(1/\rho_{xx})(d\rho_{xx}/dP) = \xi(x_{MI} - x)^{-1}(dx_{MI}/dP) < 0$, which is experimentally consistent with $dx_{MI}/dP$

< 0 (Figs. 7(b) and 9(c)). Therefore, the metallic region is expanded by the application of external pressure; that is, the band energy increases relative to the interactions between the conduction holes and the effect of disorder, and this increase results in carrier delocalization. Interestingly, for $x = -0.02$, the sign of $(1/\rho_{xx})(d\rho_{xx}/dP)$ changes from negative to positive at $P = 1.2$ GPa (Fig. 9(b)). This change suggests that the pressure coefficient of $\rho_{xx}$ is a measure of proximity to the critical points. The pressure dependence of $\rho_{xx}(T)$, $d\rho_{xx}/dP < 0$, at low temperatures in the insulator region is consistent with that observed in $Fe_2V_{1-x}Nb_xAl$, the volume of which increases as the Nb concentration increases [29].

Compared to electrons doped as a result of excess Fe atoms, holes doped as a result of excess V atoms have a localized nature, as indicated by comprehensive macroscopic measurements at $x < 0$ [7, 11–13]. The considerable enhancements of $\rho_{xx}$ and $C/T$ at low temperatures and the emergence of the isolated spins are reflected by the presence of a MI quantum phase transition at $x \approx -0.05$. The existence of the MI quantum critical point seems to be supported by the fact that $\Delta C/T$ at low temperature diverges with decreasing temperature, i.e., $C/T \approx T^{\alpha-1}$ $(1 > \alpha > 0)$.

To confirm the MI transition and determine its mechanism, we investigated the temperature variation of conductivity ($\sigma_{xx} = 1/\rho_{xx}$) at low temperatures under a magnetic field in the vicinity of the MI boundary. The metallic state of the perovskite $LaCo_{1-y}Ni_yO_3$ with $y > 0.4$ exhibits a Mott–Anderson transition at $y = 0.4$ [30], and the temperature variation of the conductivity is given by

$$\sigma_{xx} = \sigma_0 + mT^{1/2} \tag{3}$$

where the residual (zero-temperature) conductivity $\sigma_0$ is a measure of whether the system is a metal ($\sigma_0 \neq 0$) or an insulator ($\sigma_0 = 0$). The second term, which is proportional to $T^{1/2}$, results from an electron–electron correlation, and the coefficient $m$ can be either positive or negative [28]. The $\sigma_0$ component of the conductivity at $x = -0.02$ was clearly larger than the $\sigma_0$ component at $x \leq -0.05$. At $x = -0.05$, variable-range hopping conduction was indicated below $T \approx 40$ K [14, 15], whereas at $T < 2$–4 K, the $\sigma_{xx}(T)$ curve deviated clearly from the curve for variable-range hopping conduction (Fig. 10(a)) and seemed to follow eq. (3) at low temperatures. Equation (3) accurately represents $\sigma_{xx}(T)$, and the zero-temperature conductivity $\sigma_0$ obtained by extrapolation to $T = 0$ was zero within experimental error at $H = 0$. A finite value of $\sigma_0$ was induced by application of a magnetic field, whereas $m$ was nearly independent of $H$ (Fig. 10(a)–(b)). That is, a field-induced MI transition occurred for $x = -0.05$ but not for $x = -0.10$ up to $H = 120$ kOe (Fig. 10(b), inset).

In the $-0.02 \leq x \leq 0.10$ region, it is fair to say that the zero-temperature conductivity $\sigma_0$ can be estimated by using eq. (3) at low temperatures (Fig. 11(a)). The steep variation of $\sigma_0$ between $x = -0.05$ and $-0.02$ shown in Fig. 11(b) might be a sign of a MI transition. Therefore, it is worthwhile to investigate the pressure dependence of $\sigma_{xx}(T)$ for $x = -0.05$. Figure 12(a) shows clearly that the temperature dependence of the conductivity followed eq. (3) below 6 K. Remarkably, $\sigma_0$ emerged when pressure was applied, and $\sigma_0$ scaled approximately as $(P - P_c)^\gamma$ with a critical pressure $P_c$ of $0.56 \pm 0.09$ GPa and a critical exponent $\gamma$ of $0.54 \pm 0.13$. The application of pressure facilitated delocalization of electrons (holes), and a MI transition occurred at $P_c = 0.54$ GPa. The $T^{1/2}$ coefficient $m$ was approximately 25 $\Omega^{-1}$ cm$^{-1}$ K$^{-1/2}$, which is larger than the coefficients generally observed in the range of $-10 < m < 10$ $\Omega^{-1}$ cm$^{-1}$ K$^{-1/2}$; and $m$ was independent of pressure for $x = -0.05$ (Fig. 12(b)). At high pressure, $m$ was enhanced near the QCPs at $x \approx -0.05$ and 0.05. In contrast, $m$ showed no anomalous behavior in the experimental pressure range shown in Fig. 12(b). These facts suggest that the Griffiths–McCoy singularities, $\Delta C/T \approx \rho_{xx} \approx T^{\alpha-1}$, enhanced around the critical points might overlap with the perturbative quantum correlation in resistivity in disordered metals, $\delta\sigma_{\text{int}} = mT^{1/2}$.

Next we comment on the possible origin of the MI transition at $x_{\text{MI}} \approx -0.05$. Ślebarski *et al.* comprehensively investigated the hole-doped system $Fe_2V_{1-y}Ti_yAl$, which exhibits a paramagnetic-to-ferromagnetic transition at $y \approx 0.1$ [16]. Their theoretical analysis indicated that a heavy-hole state arises just below the Fermi level, and they claimed that at $y < 0.1$, the insulating state is stabilized by the creation of a hybridization gap at the Fermi level and, hence, that a Kondo semiconducting state is realized in $Fe_2V_{1-y}Ti_yAl$. An itinerant ferromagnetic state emerges when $y$ is increased above approximately 0.1 (correspondingly, the number of doped holes $n_c$ is approximately 0.1). In contrast, the MI transition in $Fe_{2+x}V_{1-x}Al$ seems to be quite unusual compared with that in $Fe_2V_{1-y}Ti_yAl$; that is, the nonmagnetic insulating state of the former is stabilized down to $x = -0.10$ (the number of doped holes $n_c$ is approximately equal to $3x$, or 0.3. The antisite vanadium atom at the Fe site acts as a strong scattering center and brings the conduction carriers under the strong influence of disorder. This system is probably highly resistive, having a mean free path $\lambda_e$ comparable to the lattice constant and the Fermi wavelength $\lambda_F$. To distinguish metallic and insulating states of certain materials, the so-called Ioffe–Regel condition can generally be applied. That is, metallic conduction is realized when $k_F\lambda_e > 1$, where $k_F$ is the Fermi wave vector; whereas when $k_F\lambda_e < 1$, the system becomes insulating. The characteristic Fermi wave vector and Fermi wavelengths are given by the following equations, respectively: $k_F = 2\pi/\lambda_F = (3\pi^2 n_c/M)^{1/3}$ and $\lambda_e = (\hbar/e^2)\sigma(3\pi^2/n_c^2 M)^{1/3}$, as a function of $n_c$ and $\sigma_{xx}$ obtained experimentally in this work; $M$ is the degeneracy of the conduction band maximum [31]. Assuming that $M = 1$ for $Fe_{2+x}V_xAl$, the values of $k_F\lambda_e$, a measure of the conduction state, were estimated to be 5.84, 0.99, and 0.64 for $x = 0.10$, $-0.05$, and $-0.10$, respectively. Also, the obtained value of $k_F\lambda_e$ validated roughly that the MI transition occurred at $x \approx -0.05$. Note that the Ioffe–Regel condition can be applied independently of the mechanism of the MI transition. On the basis of the small values of mobility $\mu$ ($= \sigma/en_c$), which were estimated to range from approximately $10^{-2}$ to $10^{-3}$ cm$^2$ V$^{-1}$ s$^{-1}$ for $x = -0.05$ and $-0.10$, it is plausible that the excess of vanadium atoms imparts strong disorder accompanied by electron–electron correlation. Therefore, we characterized the MI transition as arising from the interplay between disorder and correlation; that is, a Mott–Anderson transition might occur at $x \approx -0.05$.

*3.5 Quantum critical singularities in Fe$_{2+x}$V$_{1-x}$Al*

Since Nishino *et al*. [11] first reported the enhancement of $C/T$ and the negative TCR in Fe$_2$VAl, distinctively different theoretical explanations of these observed properties have been published. Specifically, Weht and Pickett [4] proposed an excitonic correlation in the compensated semimetallic band structure, and Singh and Mazin [5] reported a spin fluctuation in the vicinity of the paramagnetic-to-ferromagnetic quantum transition. Experiments, along with the latter, showed that the strong specific heat enhancement originates from ferromagnetic spin fluctuation in Fe$_2$VAl having the antisite defects and in the small off-stoichiometric Fe$_{2+x}$V$_{1-x}$Al. In fact, the ferromagnetic QCP is located at $x \approx 0.05$ and not at the stoichiometric composition ($x = 0$) [12]. Additionally, in this paper, we demonstrated that the MI quantum phase transition takes place at $x \approx -0.05$. The MI transition seems to be induced both by electron–electron correlation and by disorder but not by excitonic (electron–hole) correlation, as indicated by a theoretical consideration [4]. On the basis of our findings, we propose the schematic *x–T* phase diagram shown in Fig. 13. Our findings provide important insight into the singularities observed in Fe$_2$VAl. In the Heusler (L2$_1$) structure, several different atomic disorders types are known [10]. Introduction of such disorders in Fe$_2$VAl modifies the band structure, and as a result, the electronic structure is modified considerably [32]. Following the classification of the disorders in Refs. 10 and 32, we note the defect created mainly in $Fe_{2+x}V_{1-x}Al$ as a $BiF_3$-type ($DO_3$) disorder. The antisite defects associated with the Fe and V sites, the $BiF_3$-type disorder, create a local heterogeneous state consisting of quite different electronic states compared with the non-magnetic semimetallic state in a defect free $Fe_2VAl$, for example, a ferromagnetic cluster with a delocalized character and a localized hole. The origins of the enhancements in $C/T$ and $S$ and the TCR observed at high temperatures are still controversial but can be understood as being consequences of the existence of the QCPs close to each other in $Fe_{2+x}V_{1-x}Al$, and the accompanying charge and magnetic fluctuations. Interestingly, both of the leading theories [4, 5]—which try to explain the mass enhancement in $Fe_2VAl$ by quite different mechanisms, that is, charge and ferromagnetic fluctuations, respectively—might have to be reconsidered. Furthermore, because there are many kinds of antisite defects in Heusler structures [10, 32], we can expect a wide variety of emergent phenomena resulting from the disorders introduced as a result of antisite defects, elemental substitutions and the off-stoichiometries, *e.g.*, $Fe_{2-x}VAl_{1+x}$ [33] and $Fe_2VAl_{1-\delta}$ [7, 34] in addition to $Fe_{2+x}V_{1-x}Al$.

**Summary**

We confirmed that in the Heusler-type compound $Fe_{2+x}V_{1-x}Al$, a MI quantum phase transition occurs at $x \approx -0.05$. Under high pressure, a finite zero-temperature conductivity $\sigma_0$ appears and follows a power law, $\sigma_0 \approx (P - P_c)^{\gamma}$, with a critical pressure $P_c$ of $0.56 \pm 0.09$ GPa and a critical exponent $\gamma$ of $0.54 \pm 0.13$, respectively, for $x = -0.05$. Likewise, application of a magnetic field facilitates the emergence of finite conductivity at $T = 0$ K. Below $x = 0$, a paramagnetic spin state appears with localized futures; whereas above $x = 0$, a number of spins interact with each other with a ferromagnetic correlation, and, as a result, itinerant ferromagnetism emerges above $x = 0.05$ [12].

**Acknowledgements**

One of the authors (TN) acknowledges NWO (Dutch Organization for Scientific Research) for a visitor grant.

**Figures and Captions**

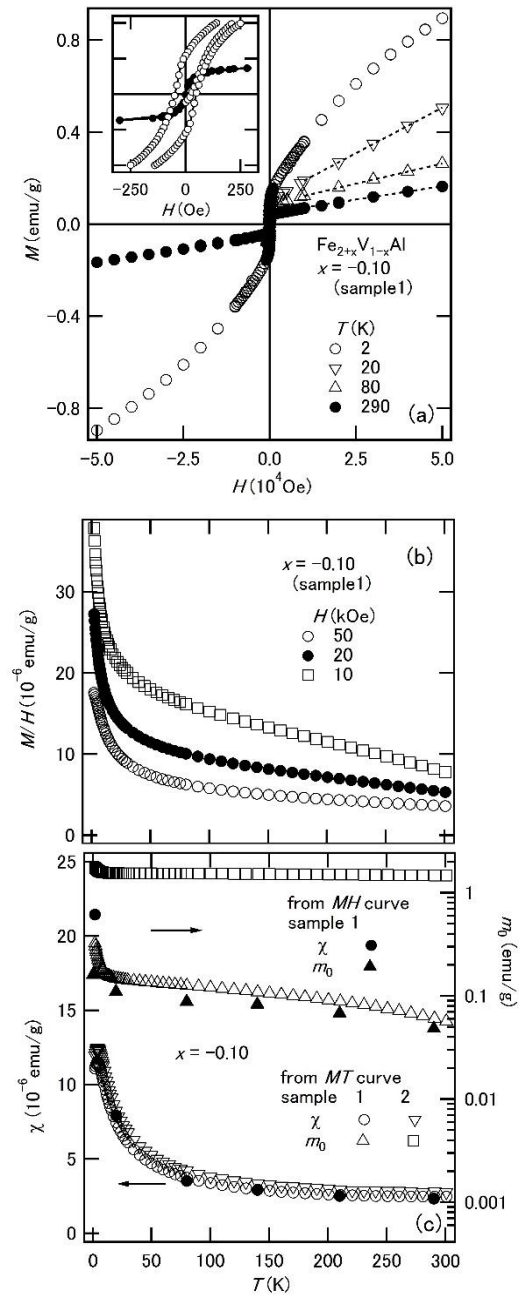

FIG. 1 (a) Magnetization–field (*MH*) curves at various temperatures for $x = -0.10$ (sample 1). The inset shows an expanded view of the *MH* curves at $T = 2$ and 290 K. (b) Temperature dependence of *M*/*H* measured at various fields. (c) Linear and constant components of *M* with respect to *H*, $\chi$, and $m_0$, as a function of temperature. Note that $\chi$ and $m_0$ obtained from the *MH* curves (solid symbols) correspond well with those obtained from the *MT* curves (open symbols).

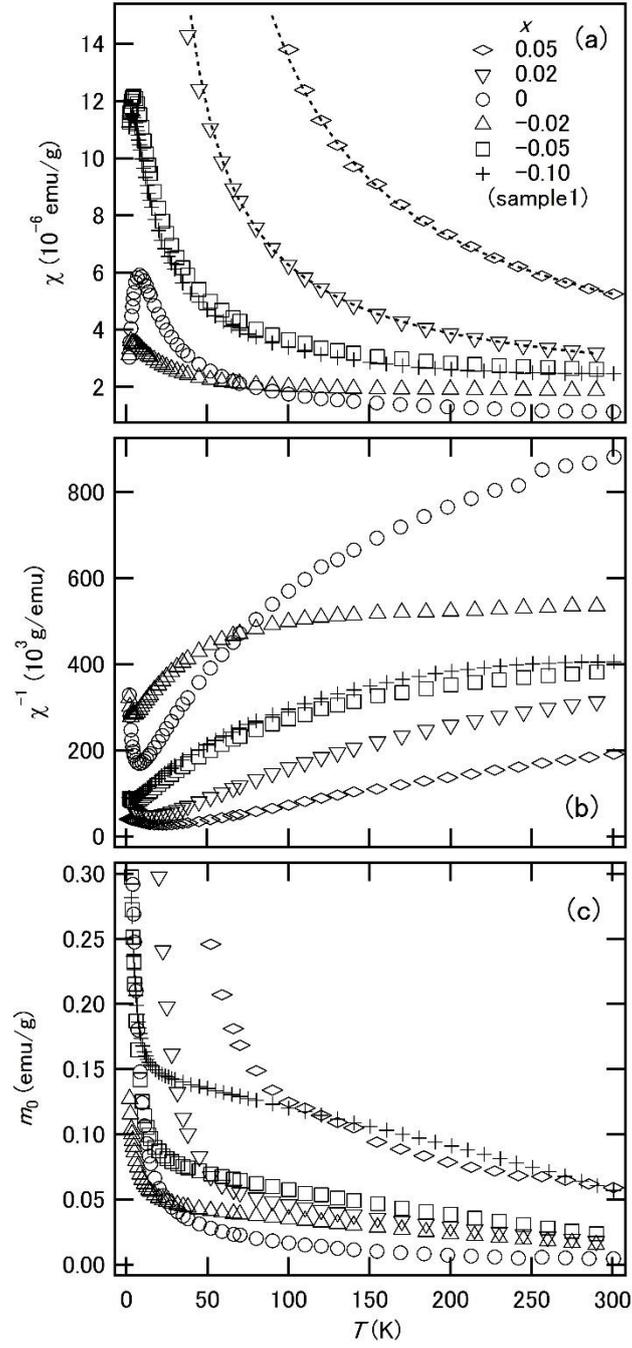

FIG. 2 Temperature dependences of (a) $\chi$, (b) $\chi^{-1}$, and (c) $m_0$ at various values of $x$.

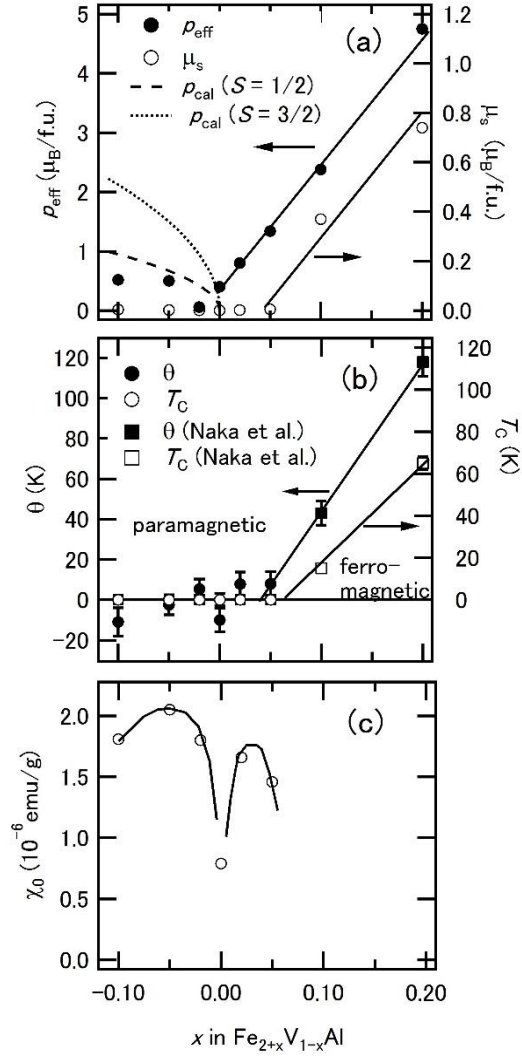

FIG. 3 (a) Variations of $p_{eff}$ and $\mu_s$ with $x$, measured at $T = 2$ K; $p_{eff}$ was obtained by means of a modified Curie–Weiss law. Dashed and dotted curves represent the calculated values based on a localized spin model with $S = 1/2$ and $3/2$, respectively (see the text for details). (b) Weiss temperature $\theta$ (this work and Naka *et al.* [12] for $x \geq 1.0$) and Curie point $T_C$ [12] as a function of $x$. (c) Constant susceptibility $\chi_0$ as a function of $x$. Solid lines are aids for visualization.

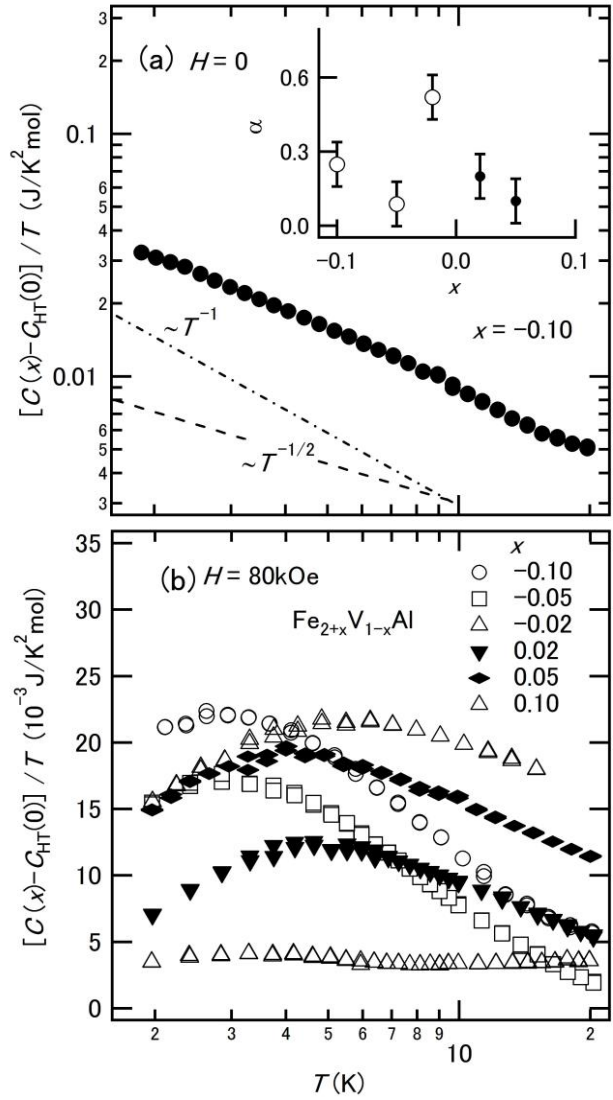

FIG. 4 (a) Double logarithmic plot of $[C(x) - C_{HT}(0)]/T$ below $T = 20$ K for $x = -0.10$. The inset shows the $x$ variation of the exponent $\alpha$ determined from $[C(x) - C_{HT}(0)]/T \approx T^{\alpha-1}$. Open and solid symbols represents the values obtained by this work and by Naka et al. [21], respectively. (b) $[C(x) - C_{HT}(0)]/T$ as a function of temperature at $H = 80$ kOe for $-0.10 \leq x \leq 0.10$.

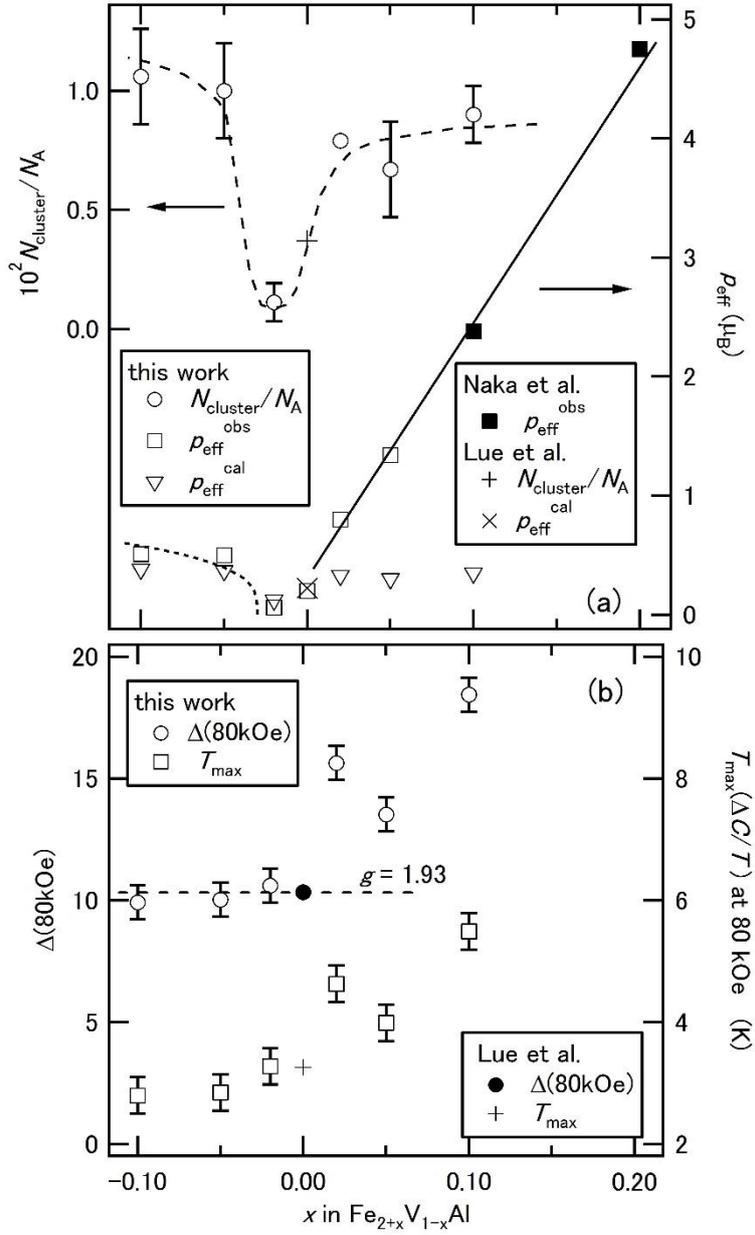

FIG. 5 (a) The $x$ dependences of the number of clusters $N_{\text{cluster}}$ normalized by the Avogadro constant $N_A$ and the calculated effective moment $p_{\text{eff}}$(cluster) estimated by fitting of the field-induced anomaly to the multilevel Schottky model with the assumption of $J = 3/2$ at $H = 80$ kOe and (b) the energy gap $\Delta$ and the peak temperature $T_{\max}(\Delta C/T)$ for $-0.10 \leq x \leq 0.10$. Here $\Delta C/T$ is defined to be $[C(x) - C_{\text{HT}}(0)]/T$. For comparison, the observed effective moment $p_{\text{eff}}$(observed) is also plotted in panel (a). All dashed and solid lines are aids for visualization.

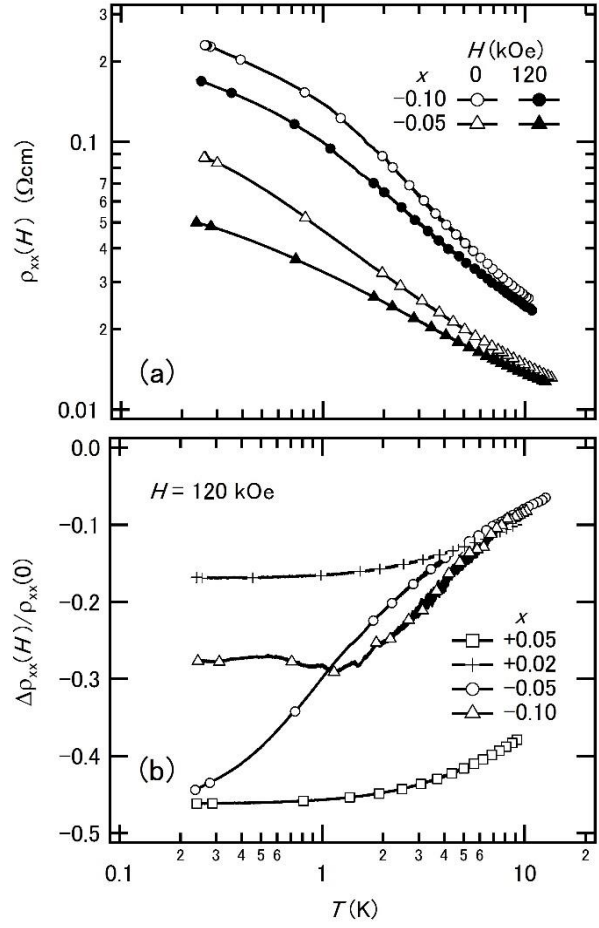

FIG. 6 Temperature dependences of (a) $\rho_{xx}(H)$ for $x = -0.10$ (sample 1) and $-0.05$ measured at $H = 0$ and 120 kOe and (b) $\Delta\rho_{xx}/\rho_{xx}(0)$ $(= [\rho_{xx}(H) - \rho_{xx}(0)]/\rho_{xx}(0))$ for $-0.10 \leq x \leq 0.05$.

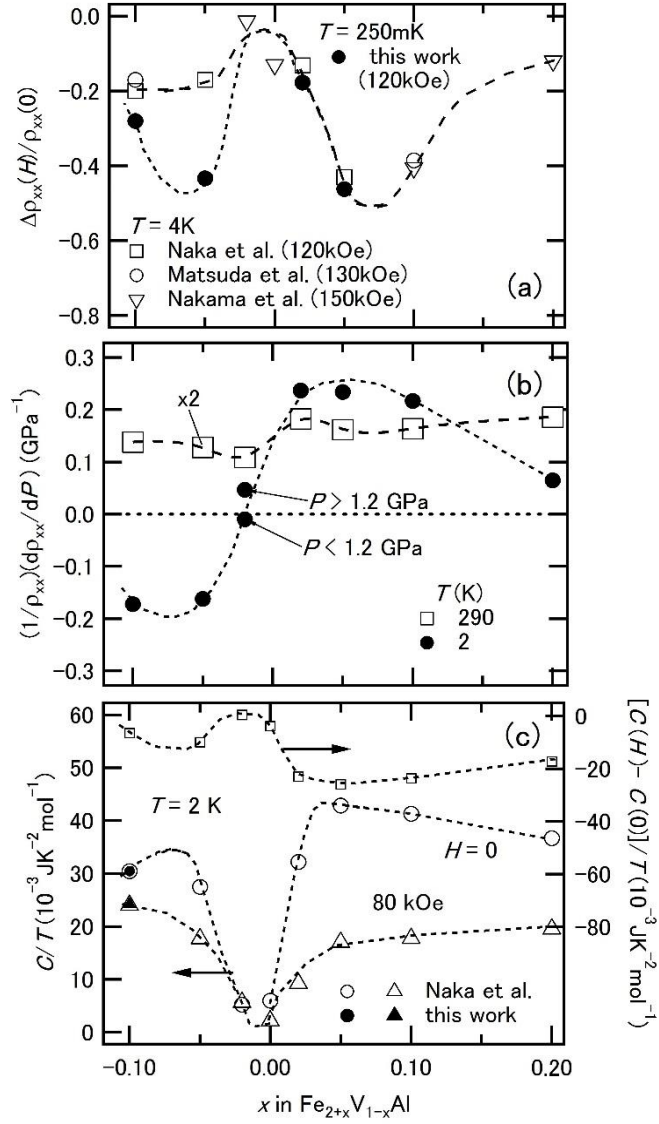

FIG. 7 (a) Variation of magnetoresistance ratio $\Delta\rho/\rho$ at $T = 4$ and 0.25 K as a function of $x$. Data obtained at $T = 4$ K are taken from the literature by Naka *et al.* [12], Matsuda *et al.* [14], and Nakama *et al.* [13]. (b) Pressure coefficient of $\rho_{xx}$ measured at $T = 2$ and 290 K as a function of $x$. (c) Variations of $C/T$ measured at $T = 2$ K under an external magnetic field of $H = 0$ and 80 kOe as a function of $x$. Note that $C/T$ was suppressed strongly by the applied magnetic field, especially, at $x \approx 0.05$ and $-0.05$. Data obtained at $x \geq -0.05$ are taken from Ref. 12. The dashed and solid lines are aids for visualization.

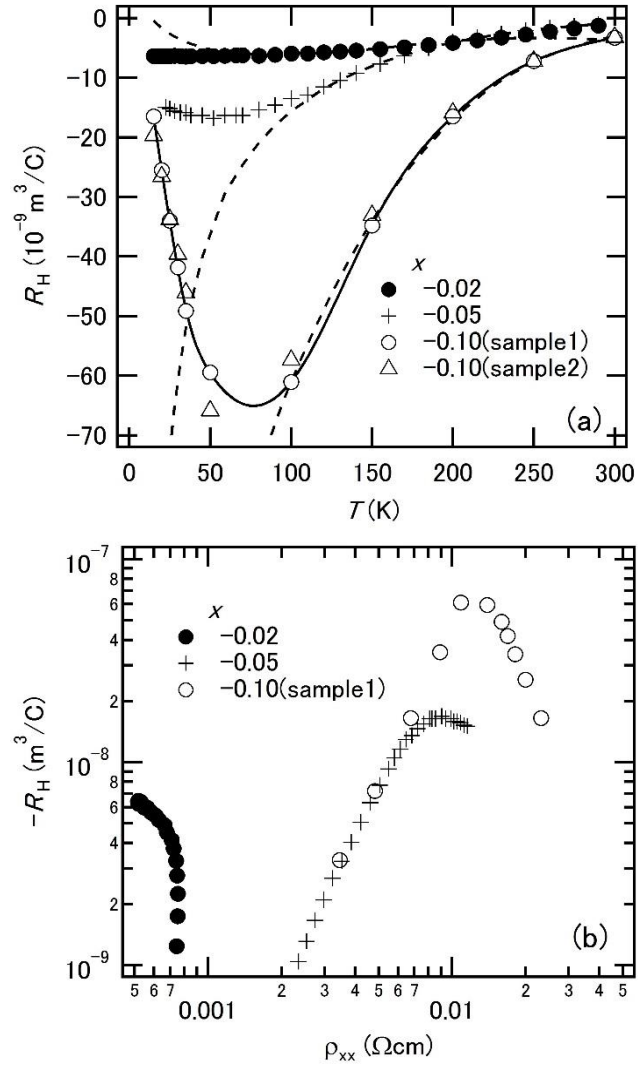

FIG. 8 Hall coefficient as a function of (a) temperature and (b) resistivity for $x = -0.10$, $-0.05$, and $-0.02$. Dashed lines are fits to eq. (2). Solid lines are aids for visualization. The Hall coefficients for $x = -0.05$ and $-0.02$ are taken from the literature [8].

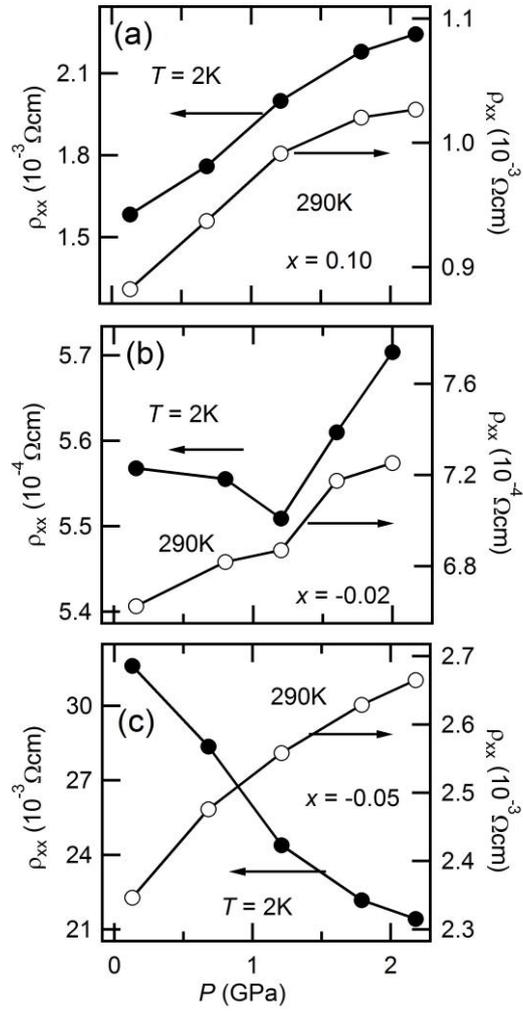

FIG. 9 Pressure dependences of $\rho_{xx}$ measured at $T = 2$ and 290 K for $x = 0.10$, $-0.02$ and $-0.10$, respectively.

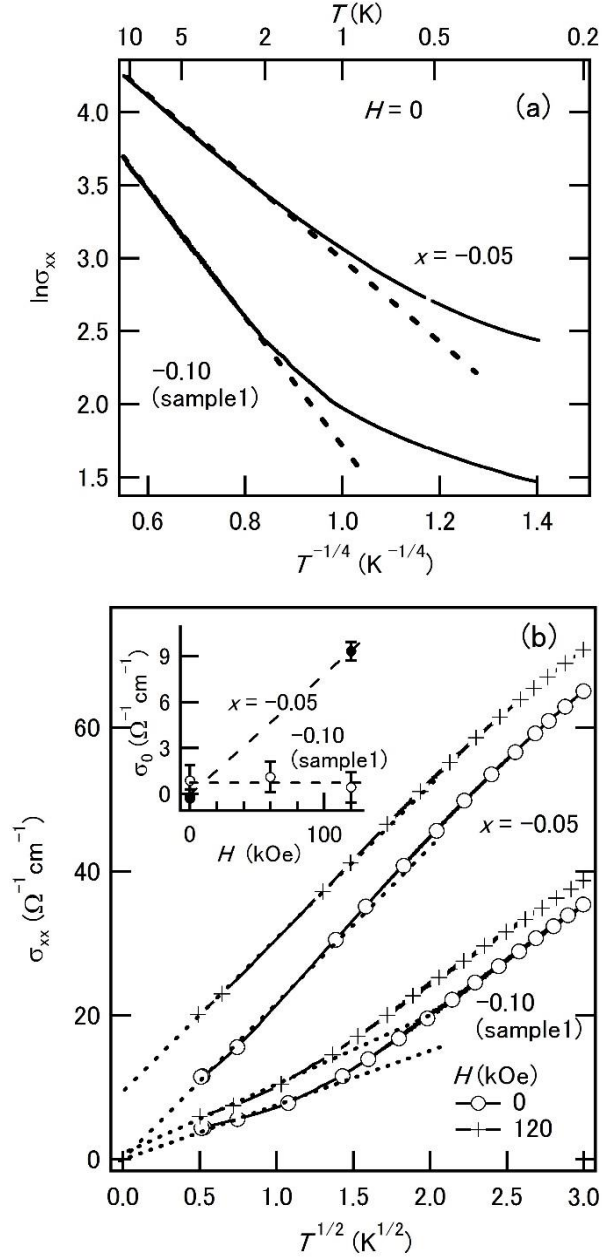

FIG. 10 (a) Logarithm of $\sigma_{xx}$ as a function of $T^{-1/4}$ for $x = -0.10$ (sample 1) and $-0.05$ measured at $H = 0$. Dashed lines represent the temperature dependence established for variable-range hopping conduction at temperatures above approximately 2 K. (b) Conductivity as a function of $T^{1/2}$ for $x = -0.10$ (sample 1) and $-0.05$ measured at $H = 0$ and 120 kOe. The inset shows the field dependence of $\sigma_0$ for $x = -0.10$ (sample 1) and $-0.05$. Dotted lines are aids for visualization.

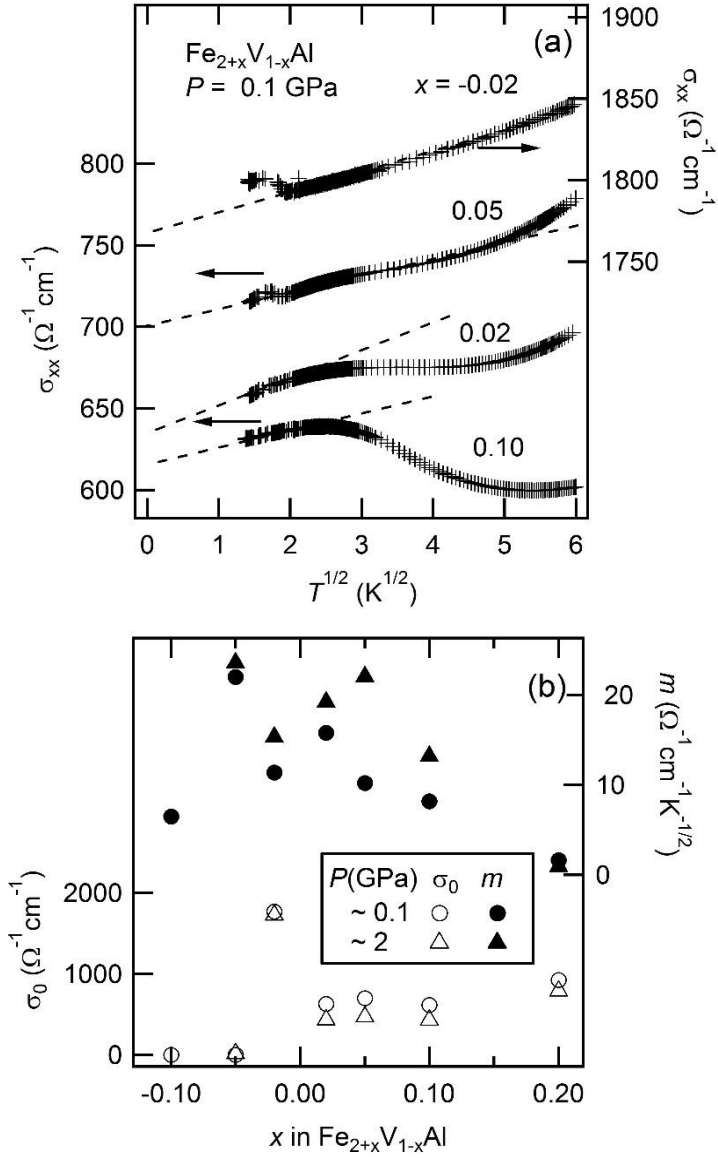

FIG. 11 (a) Variations of $\sigma_{xx}$ as a function of $T^{1/2}$ for $x = -0.10, -0.05, -0.02, 0.02, 0.05$, and $0.10$. (b) Variations of $\sigma_0$ and $m$ with $x$ at $P \approx 0.1$ and 2 GPa. The data is taken from Ref. 15 for $x = -0.05, 0.02$ and $0.10$.

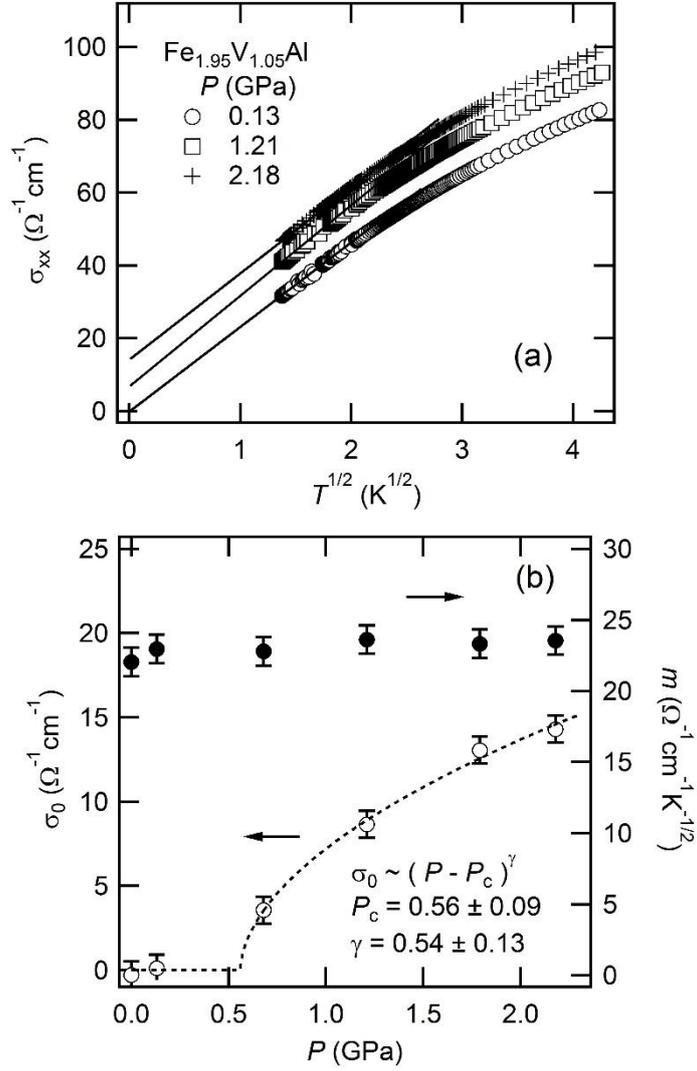

FIG. 12 (a) Variations of $\sigma_{xx}$ as a function of $T^{1/2}$ for $x = -0.05$ at various pressures. The solid lines show fits to $\sigma_{xx} = \sigma_0 + mT^{1/2}$. (b) Variations of $\sigma_0$ and $m$ as a function of pressure. The dashed line represents a power law, $\sigma_0 \approx (P - P_c)^\gamma$.

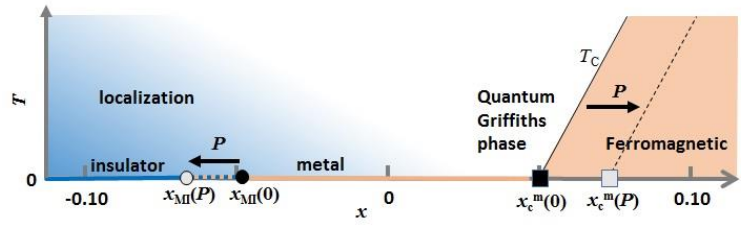

FIG. 13 Schematic $x$–$T$ phase diagram of $Fe_{2+x}V_{1-x}Al$. The Curie point $T_C$ emerges above the ferromagnetic-paramagnetic QCP $x_c^m$ and increases with increasing $x$. $T_C(x)$ line shifts toward higher $x$ under pressure, $i.e.$, $x_c^m$ increases. The MI transition point $x_{MI}$ locates at $T = 0$ K and decreases with increasing pressure.